\documentstyle[11pt,aaspp4]{article}

\slugcomment{Accepted for August 2003 A.J.}
\lefthead{Harris et al.}
\righthead{White Dwarfs in the SDSS}

\begin{document}

\title{An Initial Survey of White Dwarfs in the Sloan Digital Sky Survey}

\author{Hugh C. Harris\footnote{
  U.S. Naval Observatory, PO Box 1149, Flagstaff, AZ 86002-1149.},
James Liebert\footnote{
  Steward Observatory, University of Arizona, 933 N. Cherry Ave.,
  Tucson, AZ 85721-0065.},
S. J. Kleinman\footnote{
  Apache Point Observatory, PO Box 59, Sunspot, NM 88349-0059.},
Atsuko Nitta$^3$,
Scott F. Anderson\footnote{
  Dept. of Astronomy, University of Washington, Box 351580,
  Seattle, WA 98195-1580.},
Gillian R. Knapp\footnote{
  Princeton University Observatory, Peyton Hall, Princeton, NJ 08544-1001.},
Jurek Krzesinski$^{3,}$\footnote{
  Obserwatorium Astronomiczne na Suhorze, Akademia Pedagogicazna w Krakowie,
  ulica Podchor\c{a}\.{z}ych 2, PL-30-084 Krak\'{o}w, Poland.},
Gary Schmidt$^2$,
Michael A. Strauss$^5$,
Dan Vanden Berk\footnote{
  Dept. of Physics and Astronomy, University of Pittsburgh,
  3941 O'Hara Street, Pittsburgh, PA 15260.},
Daniel Eisenstein$^2$,
Suzanne Hawley$^4$,
Bruce Margon\footnote{
  Space Telescope Science Institute, 3700 San Martin Drive, Baltimore,
  MD 21218.},
Jeffrey A. Munn$^1$,
Nicole M. Silvestri$^4$,
Allyn Smith\footnote{
  Dept. of Physics and Astronomy, University of Wyoming, P.O. Box 3905,
  Laramie, WY 82071.}.
Paula Szkody$^4$,
Matthew J. Collinge$^5$,
Conard C. Dahn$^1$,
Xiaohui Fan$^2$,
Patrick B. Hall$^{5,}$\footnote{
  Departamento de Astronom\'{\i}a y Astrof\'{\i}sica, Facultad
  de F\'{\i}sica, Pontificia Universidad Cat\'{o}lica de Chile,
  Casilla 306, Santiago 22, Chile.},
Donald P. Schneider\footnote{
  Dept. of Astronomy and Astrophysics, The Pennsylvania State University,
  525 Davey Laboratory, University Park, PA 16802.},
J. Brinkmann$^3$,
Scott Burles\footnote{
  Physics Department, Massachusetts Institute of Technology,
  77 Massachusetts Avenue, Cambridge, MA 02139.},
James E. Gunn$^5$,
Gregory S. Hennessy\footnote{
  U.S. Naval Observatory, 3450 Massachusetts Avenue, NW,
  Washington, DC 20392-5420.},
Robert Hindsley\footnote{
  NRL/Remote Sensing Division, Code 7210, Naval Research Laboratory,
  4555 Overlook Avenue, SW, Washington, DC 20375.},
Zeljko Ivezi\'c$^5$,
Stephen Kent\footnote{
  Fermi National Accelerator Laboratory, P.O. Box 500, Batavia,
  IL 60510.}$^,$\footnote{
  Department of Astronomy and Astrophysics, The University of Chicago,
  5640 South Ellis Avenue, Chicago, IL 60637.},
Donald Q. Lamb$^{16}$,
Robert H. Lupton$^5$,
R. C. Nichol\footnote{
  Dept. of Physics, Carnegie Mellon University, 5000 Forbes Avenue,
  Pittsburgh, PA 15232.},
Jeffrey R. Pier$^1$,
David J. Schlegel$^5$,
Mark SubbaRao$^{16}$,
Alan Uomoto\footnote{
  Dept. of Physics and Astronomy, The Johns Hopkins University,
  3400 North Charles Street, Baltimore, MD 21218-2686.},
Brian Yanny$^{15}$,
Donald G. York$^{16}$}

\begin{abstract}
An initial assessment is made of white dwarf and hot subdwarf stars
observed in the Sloan Digital Sky Survey.  In a small area of sky
(190 square degrees), observed much like the full survey will be,
269 white dwarfs and 56 hot subdwarfs are identified spectroscopically
where only 44 white dwarfs and 5 hot subdwarfs were known previously.
Most are ordinary DA (hydrogen atmosphere) and DB (helium) types.
In addition, in the full survey to date, a number of WDs have been
found with uncommon spectral types.  Among these are blue DQ stars
displaying lines of atomic carbon; red DQ stars showing molecular
bands of C$_2$ with a wide variety of strengths; DZ stars where Ca
and occasionally Mg, Na, and/or Fe lines are detected; and
magnetic WDs with a wide range of magnetic field strengths in
DA, DB, DQ, and (probably) DZ spectral types.
Photometry alone allows identification of stars hotter than
12000~K, and the density of these stars for $15<g<20$ is found
to be $\sim$2.2~deg$^{-2}$ at Galactic latitudes 29--62$^\circ$.
Spectra are obtained for roughly half of these hot stars.
The spectra show that, for $15<g<17$, 40\% of hot stars are WDs 
and the fraction of WDs rises to $\sim$90\% at $g=20$.
The remainder are hot sdB and sdO stars.
\end{abstract}

\keywords{Stars: chemically peculiar --- Stars: magnetic fields
          --- Surveys --- White Dwarfs}

\section{Introduction}

There are currently 2249 known white dwarfs (WDs) with spectroscopic
confirmation in the McCook \& Sion (1999) catalog.
These have been found in a variety of ways, but primarily the
hotter stars are found as UV-excess or blue objects in imaging surveys,
while the cooler stars are found in proper motion surveys.
Some fairly homogeneous surveys have been published of hotter
blue stars, such as the Palomar Green survey (PG: Green, Schmidt
\& Liebert 1986), the Kiso survey (Darling \& Wegner 1996),
and the more recent Hamburg/EOS surveys (Homeier et al.\ 1998;
Christlieb et al.\ 2001; Koester et al.\ 2001).  In recent years the
Luyten Half Second Catalog (LHS: Luyten 1979) has been a prime source of
cooler objects.  Although largely complete to magnitude and color limits
over 10,000 square degrees at high Galactic latitudes, the PG found 
fewer than 400 hot WDs to a blue magnitude limit just fainter
than $B = 16$.  Fleming, Liebert \& Green (1986) analyzed the PG sample in
a first comprehensive estimate of the formation rate and luminosity
function of hot WDs.  Other estimates, however, have suggested
a somewhat higher space density of these objects, while recent estimates
of the formation rate of planetary nebulae exceed the inferred formation
rate of WDs (Pottasch 1996, and references therein).

Deeper imaging surveys are necessary to sample the disk scale height
for WDs adequately.  Boyle (1989) found a value of 275$\pm$50 pc,
based on a sample of 41 stars to a magnitude limit of B=20.9.  Much larger
imaging surveys to fainter magnitudes than PG have been underway for
several years, but only limited samples with spectroscopic confirmations
have been published so far.  Like the PG, the point sources in these
surveys are measured with only coarse photographic magnitudes or grism
spectra and cover only a few color bands.

The Sloan Digital Sky Survey (SDSS: York et al. 2000)
is obtaining deep CCD images of 10,000 deg$^2$ of the north
Galactic cap, overlapping most of the PG but going much deeper, and in 
five photometric bands.  The SDSS bands ($u$, $g$, $r$, $i$, and $z$)
cover the entire optical range from the UV atmospheric cutoff (3200\AA)
to the detector's red sensitivity cutoff ($\sim$10,000\AA) --- see
Fukugita et al. (1996) for details of the photometric system. 

It was apparent from the detailed simulations of Fan (1999) that the
SDSS would be very efficient in finding WDs, especially the hot
objects detectable by UV-excess.  This study suggested that something like
30,000 WDs would be detected to $r < 19.5$.  Fan (1999) showed
that hot WDs would be separable for the most part from quasars,
compact emission-line galaxies, and other stars using their locations in
the SDSS three-color (ugri) diagram.  However, the WDs cooler
than about 12,000~K generally would blend in with the colors of QSOs,
and cooler than about 7000~K would blend with main-sequence stars.
We note, however, that cool WDs with unusual spectra ---
energy distributions blanketed by unusually strong absorption lines or
bands --- could be expected to be found at different locations in the
three-color diagram (Fig. 17 of Fan 1999).  

In this paper we report on the WDs and other hot stars found in one
region of sky (190~deg$^2$, referred to below as the 41-plate sample),
as well as some WDs of special interest found elsewhere.
While this sampling is in no way complete, it is shown
that the basic hot DA (hydrogen atmosphere) and non-DA (helium
atmosphere) sequences are identifiable, that they are found in about
the numbers predicted by the \cite{fan99} simulations, and that rare
kinds of cooler WDs are detected as well.
In Section 2 we discuss in more detail the color selection.
Spectra of several different types are presented in Section 3.
The final section assesses what we have learned about the potential
of the SDSS from this first examination of WD candidates.

\section{Color selection}

The colors of WDs and other objects in the SDSS photometric system
have been predicted (Fan 1999; Lenz et al. 1998) by synthesizing
their colors from model atmospheres and from observed spectra.
Further information is available from observed colors
(Krisciunus et al.\ 1998; Finlator et al.\ 2000).  A conclusion from
these studies is that the hot end of the WD sequence has colors separated
in SDSS color space from almost all other objects except hot subdwarfs.
At temperatures cooler than about 12000~K, the WD sequence begins
overlapping the colors of low-redshift QSOs and emission-line galaxies.

Objects are selected for SDSS spectroscopic observation according to
numerous criteria based on each object's color and image morphology.
High priority is given to QSO candidates chosen with colors outside the
stellar locus (Richards et al.\ 2002),
some of which turn out to be WDs.  A small fraction of fibers
remaining after galaxy, QSO, and calibration selections have been made
are assigned to candidates for different types of stars.
Present categories include (1) horizontal-branch stars,
(2) very cool stars, (3) cool subdwarfs, (4) carbon stars,
(5) WDs with unusual colors, (6) cataclysmic variables,
and (7) other objects with unusual colors.
Stars in the last three categories, plus some hot stars chosen for
calibration, often prove to be WDs, and are the subject of this paper.

Sky coverage in the early SDSS imaging data (Gunn et al.\ 1998) included
most of a stripe with $-1.25^{\rm o} < {\rm{Dec}} < 1.25^{\rm o}$ and
outside of the Galactic plane.  Early spectra have been taken covering
parts of this stripe.  The present paper examines a sample of stars
in the region with $9^h40 < {\rm{RA}} < 16^h40$ at Galactic latitudes
$29^{\rm o}-62^{\rm o}$, imaged in March 1999.
The spectra from 41 spectroscopic plates\footnote{
These are Plates 266--269, 279--285, 291--293, 295--315, 342--346,
and 348.  Most are in the SDSS Early Data Release (EDR) described by
Stoughton et al.\ 2002.} 
have been searched for WDs and hot subdwarfs, including most of
this section of the equatorial stripe (but with a few gaps)
and covering 190~deg$^2$ of sky.
This sample is referred to in this paper as the 41-plate sample.
A total of 264 WDs and 57 hot subdwarfs have been identified
spectroscopically in this region.  In addition, a number of unusual WDs
found outside this region are also described here.
These 41 plates also include six cataclysmic variables identified
by \cite{szk02} and not discussed further in this paper,
as well as 18 objects with DA+M composite spectra identified by
\cite{ray03} and included in the following tables and figures.

Figure 1 presents SDSS color-color diagrams of the region occupied by
hot stars.  The classifications from the SDSS spectra of the hot WDs
and subdwarfs are indicated in the diagram.  In addition, the colors and
magnitudes of all point sources with $15<r<20$ in a smaller region of
sky are shown.  The two curves show the colors of WD model atmospheres
(\cite{ber95}) with log~$g = 7$, 8, and 9,
kindly calculated and made available to us by Bergeron.
The colors of all objects have been dereddened by the full reddening
estimated for each line of sight (Schlegel et al.\ 1998),
although this procedure slightly overcorrects for reddening for nearby
(brighter and cooler) WDs in this sample.

Fig. 1(a) is equivalent to the well-known diagrams for WDs using
$UBV$ and $uby$ filters (e.g. Eggen \& Greenstein 1965;
Graham 1972; Wegner 1979), although rotated by 90 degrees here.
The sequence of WDs and hot subdwarfs with different temperatures
is visible, and its separation at the hot end from most contaminating
stars and QSOs is clear.  In particular, WDs hotter than 12000~K
are essentially uncontaminated by everything except hot subdwarfs.
Horizontal branch and blue straggler stars are well-separated from
all WDs by having $u-g > 0.8$.
The scatter of WDs away from the central locus near the log~$g = 8$
curve is larger here than has been seen in earlier studies.
Probably the scatter seen here has significant contributions
from random photometric errors for faint stars, from errors in the
photometric calibrations, and from overcorrection for interstellar
reddening for some stars (see next paragraph).  The scatter of WDs
in Fig. 1 is under study, pending improved reprocessing of SDSS
photometry this year.

Data for the WDs and hot subdwarfs identified spectroscopically
are given in Tables 1--3:  DA and DB WDs in Table 1, WDs with less
common spectral types in Table 2, and hot subdwarfs in Table 3.
Positions at the epoch of observation (March 1999)
are those from the SDSS Astrometric Pipeline (Pier et al.\ 2003).
Magnitudes and colors are those from version 5.3 of the
Photometric Pipeline (Lupton et al. 2003, in preparation) prepared
for the initial Data Release 1.  (Data are being reprocessed with
version 5.4 of PHOTO, and ultimately will give slightly different
results.)  The photometric calibration is based on the
SDSS standard-star system (Smith et al.\ 2002) tied to the Survey
data with the Photometric Telescope (Hogg et al.\ 2001).
The full interstellar absorption $A_{u}$ from \cite{sch98}
is given for that line of sight.  Typically, the hotter stars
will be at distances of several hundred pc, and they will have
approximately the listed absorption at these high galactic latitudes.
However, only some fraction of that absorption will apply to nearby
(bright, cooler) WDs.  Reddening can be calculated using
$E(u-g)=0.264A(u)$, $E(g-r)=0.202A(u)$, $E(r-i)=0.129A(u)$,
and $E(i-z)=0.119A(u)$;  for most stars at these high latitudes,
the reddening is small whether or not the full absorption is correct.
Proper motions are calculated using the USNO-A2.0 Catalog
(Monet et al. 1999) for the first epoch position.
For a few stars not in the USNO-A Catalog, usually because
they are too faint to be detected on both the POSS-I blue and
red plates, the first epoch position is taken from the
USNO-B1.0 Catalog (Monet et al. 2003).
The proper motions typically have errors of 6--8~mas~yr$^{-1}$ rms.
Where the motion is less than 20~mas~yr$^{-1}$, it may not be real,
and the position angle of the motion is omitted.
Spectral types are from our classification of the SDSS spectra,
as discussed in the following section.

\section{White Dwarf Types}

The spectra described in this section are taken with the two SDSS
2.5 m-telescope fiber spectrographs\footnote{
See http://www.astro.princeton.edu/BBOOK/SPECTRO/spectro.html
for a more complete description of this instrument.}.
Using fibers with a diameter of 3 arcsec, about 600 objects
are observed each exposure with a resolution of about 1800.
Exposures are typically 1 hr (four integrations of 15 min),
depending on observing conditions to reach a target signal/noise.
The spectra are extracted and calibrated automatically.
Night-sky lines are not subtracted perfectly, and residuals are often
seen at 5577, 5890/5896, and 6300 \AA and at the near-IR end.  The
wavelength calibration (to heliocentric, vacuum wavelengths) is done.
On rare occasions, a spurious jump remains in a spectrum at 5900 \AA,
where data from the blue and red detectors are combined.
Spectra are corrected for terrestrial extinction, they are flux
calibrated using spectrophotometric standards on each plate,
and finally, they are corrected for the interstellar absorption
listed in Tables 1--3.

\subsection{DA, DB and DO Stars}

For this preliminary analysis, spectral types are assigned on the
Sion et al. (1983) system by comparison with the spectrophotometric
atlas of Wesemael et al. (1993).
For DA stars that appear to fall in the region (10--20,000~K)
where DA Balmer line strengths are maximum, the $u-g$ and $g-r$
colors are also taken into account in assigning a classification.
Figure 2 shows representative spectra of DA4 stars covering a range
of magnitudes.  In this and following figures, spectra are given
in units of $f_{\nu}$, with the peak flux density normalized to one,
and with each spectrum shifted by one unit.  (The coordinates and
plate--fiber number are shown at the right below each spectrum.)
We believe that these spectra are generally accurate enough for
WD/subdwarf discrimination, and for WD classification to $\pm$1
subtype, for stars with $g < 19.5$.
For the fainter stars the estimates are generally listed with a colon.

Figure 3 shows the sequence of DA types found in SDSS spectra.
The sequence covers the range from DA1 to DA7, where the index
is the conventional number corresponding to $50,400/T_{\rm eff}$.
The sequence extends to temperatures as cool as about 7000~K,
whereupon  the WD colors begin blending with colors of main-sequence
stars so badly that WDs are no longer selected for obtaining spectra.
More accurate temperature and surface
gravity determinations require fitting of the Balmer line profiles
(e.g. Bergeron, Saffer \& Liebert 1992); this kind of fitting has
begun (Raymond et al.\ 2003; Nitta et al.\ 2003), but a discussion
is deferred until a larger and more complete sample is assembled.

The majority of the DA stars appear to have $T_{\rm eff} < 25,000$~K,
with very few stars showing lines weak enough or colors blue enough
to be hotter than 40,000~K.  Conversely, few stars have been identified
at $T_{\rm eff} < 10,000$~K, a result of the color ranges chosen to
identify QSO candidates for spectroscopic observation.
The paucity of hotter stars is at least partly a consequence of the high
Galactic latitude surveyed. It is expected that these WDs are
primarily of the disk population, but the hotter WDs fainter
than $\sim$17 mag must lie beyond the scale height of the disk. 

Examples of spectra of non-DA stars are shown in Figure~4.
The non-DA stars from the 41-plate sample in Table 1 include only one
hot DO star, the top spectrum in Fig.~4 (=HE1314+0018; Christlieb
et al.\ 2001).  Another cooler DO star is shown (the second spectrum)
and listed in Table 2.  See \cite{dre96} for a discussion of this type
of star.  The few hot non-DA stars found here are consistent
with the temperature distribution of DAs, although three hot DQ stars
are discussed below.  The third and fourth spectra show DB2 stars.
The DB gap, a temperature range 30000--45000~K with no known DB stars,
lies between the DO and DB2 stars.  Fitting the line profiles of the
second, third, and fourth spectra in Fig.~4 (and other similar stars
being found by SDSS) will determine if any of these stars encroach
on the DB gap.  However, most DB stars in this sample appear to be
confined to $T_{\rm eff} < 18,000$~K, as they do not show particularly
strong He~I lines characteristic of the hotter stars. The line strengths
are quite sensitive to temperature over this range.  The relative
weakness of the 4388\AA\ line and absence of He~II 4686\AA\
distinguishes them from helium-rich subdwarfs.  He~I lines are no longer
visible at this resolution at $T_{\rm eff}$ below about 12,000~K. 

\subsection{Magnetic White Dwarfs}

A first list of magnetic WDs in the SDSS was given by \cite{gan02}
based on the EDR.  Table 2 expands that list by including 22 magnetic
WDs, including some spectra obtained since the EDR.  Additional
candidates have been found and will be discussed in a separate paper
(Schmidt et al., 2003).  Table 2 includes eight stars
discussed by G\"ansicke et al.;  the stars SDSSJ1222+0015 and
SDSSJ2323$-$0046 are omitted here, however, pending confirmation of
their magnetic nature and H/He type.  Figures 5 and 6 show spectra
of some of the magnetic DA and DB stars, respectively.
The hydrogen stars are ordered top to bottom by increasing mean field
strength as indicated by the degree of Zeeman splitting.  The strength
of the mean surface field observed in Fig.~5 ranges from about 1~MG
for the top two spectra to about 20~MG for the tenth spectrum
to roughly 120~MG and 300~MG for the last two spectra.
The last spectrum (SDSSJ2247+1456) has features (particularly near
5800, 7000, and 8400\AA) similar to those in Grw~+70$^{\circ}$8247,
a WD with a field about 320~MG (Wickramasinghe \& Ferrario 1988).
Detailed modelling of the line profiles is necessary in order to
determine the polar field strength, the viewed inclination angle,
and the degree to which the indicated field geometry is that of a
centered dipole.  Time-resolved spectrophotometry and/or
spectropolarimetry can show whether the object is rotating.

The first and second spectra of Figure~6 are two of the simplest,
low field He~I spectra of magnetic WDs yet to be obtained.
The Zeeman triplets of He~I at 4471, 4921, 5015, 5875, and 6678\AA\
can be matched with component positions from Kemic (1974) for a
mean surface field strength of 1.9~MG for SDSSJ0142+1315 and 4.4~MG
for SDSSJ0017+0041.  Neither star shows any evidence for hydrogen.
The third spectrum (SDSSJ0333+0007) is SDSS's recovery of the known WD
HE~0330$-$0002 (Reimers et al. 1998; Schmidt et al. 2001).  The latter
authors demonstrated that the spectrum is circularly polarized, thus
confirming the suggestion by the former authors that it is a magnetic
WD.  However, the overall energy distribution suggests a
somewhat lower $T_{\rm eff}$ of 6,000--7,000~K.  Thus, while there is
apparent near-coincidence with some He~I features, the star is too
cool for significant excitation of the required lower levels.  Hence,
the absorption features remain of unknown origin, but are likely due
to trace elements in a He-rich atmosphere.

Additional likely magnetic WDs with carbon and heavy 
element features, respectively, are discussed in Sections 3.3 
and 3.4 below. 

\subsection{DQ and Peculiar DQ ``Carbon'' White Dwarfs}

The majority of cool WDs are blackbody-like, showing weak or no spectral
features, and are difficult to identify because they blend in with the
normal stellar locus in two-color diagrams.  
Exceptions include WDs showing strong trace abundances of carbon,
believed to be dredged-up from the underlying core
(Wegner \& Yackovich 1984; Pelletier et al. 1986).
Note that stellar evolutionary models of asymptotic
giant branch and pre-WDs indicate that the outer part of the
carbon-oxygen core should be essentially devoid of oxygen,
so that we would expect only carbon to be dredged up via its
upward-diffusing tail (Salaris et al. 1997).
Collectively, the WDs showing atomic and/or molecular carbon features
have been classified ``DQ.''

Warmer DQ WDs at 12000--14000~K exhibit C~I and even C~II lines
(see, for example, the review of WD spectra by
Wesemael et al. 1993).  Near 10,000~K the stars with
highest C/He abundance show both atomic C~I lines and C$_2$ ``Swan"
bands in their optical spectra.  Lower temperature stars show only
the molecular features, with five principal bandheads at 4383, 4737,
5165, 5636, and 6191\AA\ , corresponding to the (2,0), (1,0), (0,0),
(0,1), and (0,2) transitions, respectively.  Each band has a sharp
red edge to the absorption, degrading gradually to shorter wavelengths.
Until recently, these stars have been found to temperatures as low
as about 6,300~K (Bergeron, Leggett, \& Ruiz 2001).
Because of the low continuous opacity exhibited by neutral helium
(primarily He$^-$), only a little carbon is required to be detected
in WDs with He-dominated atmospheres.
A C/He abundance ratio of order 10$^{-2}$ can produce C$_2$ bands
for stars with $T_{\rm eff} \sim10,000$~K, and the required abundance
decreases with decreasing temperature, so that a ratio of 10$^{-3}$
may suffice at 6,000~K.  This opacity acts to depress the SDSS $g$
flux (and, to a lesser extent, the $r$ flux), thus reddening the
$g-r$ color.  It also acts to redden the $B-V$ color (Schmidt
et al.\ 1999), and can have a dramatic effect on broadband colors.

Figure 7 shows the sequence for DQ stars with relatively weak carbon
features.  The spectra are arranged in order of $u-i$, a color
that offers both a wide wavelength baseline and two bands relatively
unaffected by carbon absorption, so that this ordering might
approximate that by (decreasing) $T_{\rm eff}$.
The top three stars in Fig.~7 show only atomic CI features;
the top star is the known DQAB star WD1728+560 (G227-5).
The detected CI lines are marked at the top of the figure.
The fourth star (SDSSJ1148$-$0126) shows both CI and molecular C$_2$ bands,
and the spectrum is nearly identical to the well-known WD0856+331
(G47-18).  The remaining stars show only weak C$_2$ bands.
Their $u-i$ colors range from 0.0 to 0.9.
Comparison with the colors of pure-He models of \cite{ber95}
suggests that these stars have $T_{\rm eff} \sim$9500--6600~K.
(We emphasize, however, that the use of pure-He models to fit the
slopes of the energy distributions is not likely to be accurate,
since important opacities are missing.  Comparison of the models
with observed colors may help us define at least a relative
$T_{\rm eff}$ scale.)

Figure 8 shows another sequence of stars with relatively strong
C$_2$ Swan bands.  Again, they are arranged in order of $u-i$ color,
covering a range 0.1--1.6.
With three exceptions (the third, fifth, and seventh stars)
the bands look fairly normal.  Their colors overlap with the weak-band
stars in Fig.~7:  over the range of 0.2--0.9 in $u-i$,
corresponding to temperatures $\sim$8600--6600~K, stars are found
with both weak and strong bands indicating a range of C/He abundances.
The colors of the last two stars are noticeably redder than the others,
and are redder than other normal DQ stars known until recently.
They imply temperatures of 5500--6000~K, although pure-He models
cannot be relied upon to give an accurate estimate.
The ninth star (SDSSJ0935+0024) has a spectrum nearly identical
to GSC2U J131147.2+292348 (Carollo et al.\ 2002; Carollo et al.\ 2003),
including two of the ultraviolet band sequence of C$_2$
at 3852 and 4102\AA; \cite{car03} derive a temperature of 5120~K
for that star.  The eighth star (SDSSJ0808+4640) has somewhat weaker
Swan bands, but with more rounded bandheads, and does not show the
ultraviolet sequence of bands.  However, despite their cool temperatures,
none of these two SDSS stars nor the GSC2 star appear to be the
``peculiar'' DQ stars discussed next.

The fifth star (SDSSJ2232$-$0744) in Fig.~8 appears somewhat different
in having bands that are more rounded and displaced toward the blue.
It may be a hybrid between the normal DQ stars and the class called
``peculiar DQ" out of considerable ignorance.  The latter begin 
appearing at temperatures below about 6300~K, with rounded, nearly
symmetrical absorption bands centered at wavelengths near
4280, 4575, 5000, 5400, and 5900\AA\ (Schmidt, Bergeron and Fegley 1995;
Schmidt et al. 1999). The former authors argue that these analogous but
apparently displaced bands (relative to those of C$_2$) are likely due
to a different but related molecule.  A detailed chemical equilibrium
analysis of H/He/C mixtures under the physical conditions encountered
in the atmospheres of these peculiar objects suggests that C$_2$H is a
molecule preferentially formed --- provided a trace abundance of
hydrogen is present.  To be sure, no calculations or measurements
exist for molecular transitions of the C$_2$H molecule.  The properties
of SDSSJ2232$-$0744, with a color-based $T_{\rm eff}$ estimate near
6,500~K (i.e. not far from the hypothesized boundary for peculiar-DQ
formation) fits nicely into this picture.  However, the existence of
three stars in Fig.~8 (as well as the GSC2 star mentioned above)
that are redder, but have apparently normal C$_2$ bands, would require
a significant range of H/He/C abundances in these stars.

The third star (SDSSJ1113+0146) and seventh star (SDSSJ1333+0016) both
exhibit distorted spectra, both have counterparts in the literature
with similar spectra, and both show significant circular polarization
in data taken at the MMT.  SDSSJ1113+0146 appears similar to LP~790-29,
a star with a temperature 7,500~K exhibiting distorted and displaced
carbon bands and strong polarization (Liebert et al.\ 1978;
Wickramasinghe \& Bessell 1979; Schmidt et al.\ 1999).
Bues (1999) modeled that star with a field strength of 50~MG.
SDSSJ1333+0016 has a spectrum with strong, scalloped carbon bands almost
identical to the spectrum of LHS~2229 (Schmidt et al. 1999).
These authors make the case that the displaced, rounded band features 
are more likely associated with those due to the hypothesized C$_2$H
bands.  Indeed, the energy distribution of LHS~2229 suggests that 
its $T_{\rm eff}$ is lower than 6300~K. 

In summary, the new DQ stars found here show a transition from stars
with atomic carbon to stars with molecular carbon that is correlated
with color (and presumably temperature) as we would expect.
However, the redder, cooler DQ stars have a wide range of C$_2$ band
strength at a given color, and the two reddest stars are probably
as cool as any DQ stars, but appear to have ``normal'' C$_2$ bands.
The interpretation is complicated by two good candidates for
highly magnetic, polarized WDs showing distorted bands.
No new detections of stars with CH features were made.

\subsection{Metallic-lined DZ Stars }

The occurrence of heavy elements --- calcium, and sometimes also
magnesium, iron, and other species --- in cool helium WD atmospheres
can result in quite strong spectral blanketing at primarily
ultraviolet wavelengths.  The origin of this additional opacity in a
small fraction of cool non-DA WDs is believed to be interstellar
or circumstellar accretion.  When the absorption is sufficiently
strong, the reddening of the $u-r$ color can pull these stars out
of the normal helium WD locus.  For this class of WD, some of the
stars found so far by SDSS substantially enrich the variety of
spectra that are known.

Most DZ stars found in SDSS spectra exhibit only the Ca~II H\&K
resonance doublet.  There are five found in the 41-plate sample
and listed in Table 2, although one (SDSSJ0944$-$0039) is hot enough
to have He lines too and is classified as DBZ.
Their spectra are shown in Figure 9, together with the brightest
of the DC stars (with no detected features) from the 41-plate sample.
That they are degenerate dwarfs and not subdwarfs is virtually
guaranteed by the nondetections of such features as the Ca~I 4227\AA\
line, the 4300\AA\ CH band, H$\beta$, the Mg~I 5175\AA\ triplet, and
the Na~I 5892\AA\ resonance doublet.  The energy distributions (flat
$g-i$ colors) suggest that these objects are warm; a Bergeron
pure-He model of 9500~K has zero color.  However, the pure-He models
may overestimate the temperatures of DZ stars (Wolff et al. 2002).
The wide range of Ca line strengths among these stars with similar
colors is obvious.  Although SDSSJ1610+0046 has weak Ca lines,
it is classified DC: as the reddening is relatively high and the
lines may be interstellar.

In Figure~10, ten spectra of DZ stars are shown that feature a detection
of metallic lines other than calcium.  Most of them show lines of
Mg~I at 3829, 3832, 3838, 5167, 5173, and 5184 \AA, and several
show the NaD lines at 5890 and 5896 \AA.
The Ca~II lines at 3933 and 3968 \AA\ are generally strong,
and the CaI lines at 4226, 4335, and 4455 \AA\ are often present.
The CaII infrared triplet at 8498, 8542, and 8662 \AA\ is clearly
present in most of the warmer stars, although the noise in the spectra
left after sky subtraction prevents seeing the lines in the fainter
stars.  The spectra are plotted in order of $r-z$ color
that is probably close to their order of temperature.
The first three stars are the warmest, with temperatures 10000--12000~K,
as indicated by their colors compared to the Bergeron models
with pure He atmospheres.
The next four spectra are redder, show generally stronger Mg lines
(including lines of MgI at 4352, 4703, and 5528 \AA, and perhaps of
MgII at 4481 \AA), and all show numerous lines of FeI --- we return
to them below.
The third and fifth spectra (SDSSJ0939+5550 and SDSSJ0956+5912)
also show Balmer lines and are DZA, an unusual type examplified
by Ross~640 (Liebert 1977).

The last three spectra in Fig.~10 have unusual (and uncertain)
classifications.  The eighth (SDSSJ0157+0033) is noisy, but has a
a strong turnover of the blue spectrum, suggesting strong Ca (and
perhaps Fe) absorption, as well as apparent complex structure
of the probable Mg~I detection.  This structure is
reminiscent of the spectrum of the only previously known magnetic DZ
star, LHS~2534 (Reid et al. 2001).  The spectrum of that star was
most noteworthy for its strong, Zeeman-split Na~I absorption,
but the weaker Mg~I lines were also split into several components.
While the S/N ratio of the spectrum of SDSSJ0157+0033 is too 
modest for a detailed comparison, the most likely interpretation
is that this star is also magnetic.  The
lack of distinct features at bluer wavelengths might also be explained
by the magnetic hypothesis.  Probable Ca~I 4227\AA\ appears unshifted
in the spectrum, as it does in LHS~2534.  A more distinct Ca~II 
absorption appears in the LHS object, but both spectra turn over 
sharply below 4500\AA.

The ninth spectrum in Fig.~10 (SDSSJ0135+1302) probably has very
strong Ca absorption and weak, barely detected Mg absorption.
It may have broad H$\alpha$ absorption, although that apparent
feature may be an artifact of incorrect flux calibration.
The tenth spectrum (SDSSJ1330+6435) appears to have extreme absorption
by NaI (and possibly by Ca, but there is insufficient flux to tell).
It appears to be very similar to that of the star 2356$-$2054 found
in the proper-motion survey of \cite{opp01}.  We find
SDSSJ1330+6435 also to have a non-zero proper motion, ruling out its
interpretation as a broad absorption line QSO.

Returning now to the fourth to seventh spectra from Fig.~10,
an expanded plot of the blue end of the spectra is shown in Figure~11,
showing details not previously seen in DZ WDs.
Detected lines of Ca, Mg, and Na are marked by open circles at the
top of Fig.~11 (and their wavelengths are listed in the text above),
and some detected lines of FeI are marked by small filled circles.
These FeI lines come from multiplet numbers
4, 15, 41, 42, 43, 45, but more lines from other multiplets are probably
present as well.  It is useful to compare these four spectra to lend
confidence that many weak features are real.  Many are also present in
the strongly blanketed DZ star G~165-7.  The FeI lines are barely
resolved (FWHM $\sim$4\AA\ in unsmoothed spectra), and Ca~I 4227\AA\ 
is also prominent and only slightly broader (FWHM $\sim$7\AA).
MgI presents itself in a variety of strengths, indicating for
SDSSJ0956+5912 and SDSSJ0937+5228 a remarkably strong MgI strength
relative to the apparent Ca~II strength.  The asymmetry in the strong
Mg~I lines due to the extended blue wings has been seen before in
G~165-7 (Hintzen \& Strittmatter 1974).  Wehrse \& Liebert (1980) argued
that the asymmetry is explained by fitting a model profile assuming a
static theory, rather than the usual impact approximation, for the van
der Waals broadening in a neutral helium-dominated atmosphere.
Traving (1968) has demonstrated such an asymmetry in laboratory
measurements.

In contrast to the fairly sharp lines of FeI and CaI in Fig.~11,
the Fe~I features in previous spectra of WDs are broader, and often
hopelessly blended.  They contribute to the reddening of the spectrum
below 4500\AA\ in the strong-DZ stars vMa~2 (Greenstein 1960) and G~165-7.
To what can we attribute the contrast between the sharp
metallic lines in Fig.~11 and the previous objects?  A temperature
difference is one possibility.  G~165-7 has an inferred $T_{\rm eff}$ of
7,500~K (Wehrse \& Liebert 1980), while several crude estimates
suggest a temperature for vMa~2 near 6,000~K.  Cool WDs have 
higher-pressure atmospheres and greater line broadening, and the 
SDSS objects could be somewhat warmer.
It has previously been suggested that cool WDs with
hydrogen-dominated atmospheres could indeed show narrow metallic lines
(Hintzen \& Strittmatter 1975).  SDSSJ0956+5912 does show up to three
Balmer lines.  The presence of hydrogen in the atmosphere may greatly
increase the opacity while decreasing the pressures where lines are
formed in the atmosphere.  Only a detailed analysis of these spectra
can show what the temperatures and abundances are.  The breadth of
the Ca~II lines and the asymmetric Mg~I profiles leave little
doubt that they are degenerate dwarfs, rather than some kind of
peculiar subdwarfs.

\subsection{Hot subdwarfs}

There are at least two spectral classes of these stars with lower gravity
than WDs (but not as low as horizontal branch stars or main-sequence
F stars), called subdwarf B (sdB) and subdwarf O (sdO).  The former
show primarily hydrogen lines, generally with helium weak or absent
(and sub-solar in abundance due to diffusion).
The generally accepted interpretation of these stars
is that they are core helium-burning stars which have
somehow been stripped of virtually all of the overlying hydrogen-rich
envelope.  They are field analogs to extreme horizontal branch (EHB)
stars found in old, especially metal-poor globular clusters.  One might
therefore expect a temperature sequence of sdB/EHB stars to link to the
blue horizontal branch (BHB) stars which occupy a well-defined, unique 
position in Sloan color-color diagrams (Yanny et al. 2000). 
However, there is a known class of helium-rich stars similar in
temperature to the sdB stars called He-sdB stars.

The 38 sdB and 18 sdO stars found in the 41-plate sample are listed in
Table 3.  They do not lie on a well-defined sequence of colors (Fig.~1),
but scatter throughout and around the colors of the hot WDs.
The spectra of the brightest eleven stars are shown in Figure 12.
Some exhibit strong helium lines in the spectra,
and are likely related to the hotter sdO or helium-rich He-sdB stars.
For a few stars here, the quality of the spectrum is too poor to
classify the object at all.  This heterogeneous group includes objects
evolved beyond the core helium-burning phase.  It is too early for a
systematic assessment of sdB, sdO, and other hot, lower gravity stars
in SDSS.  However, the substantial number listed in Table 3 shows that
the survey will ultimately allow an extensive classification and
discussion of these stars.

\section{Discussion}

Some comparison can be made between this paper and WDs discovered
previously.  A cross-check with the NLTT Catalogue (Luyten 1979) shows
reasonable consistency.  There are eight stars in Tables 1--3 that have
a proper motion $>$180~mas~yr$^{-1}$ and are bright enough ($g<19$)
that they are likely to be included in the NLTT.  All but one
(the DA4 star SDSSJ1258+0007) are in the NLTT with a proper motion
close to the motion found here.

The McCook \& Sion (1999) catalog lists 38 WDs in the region of
the 41-plate sample.
Six additional WDs in our 41-plate sample in Tables 1 and 2 have been
identified in the Hamburg/ESO Survey (Koester et al.\ 2001;
Christlieb et al.\ 2001) or in the Hamburg Quasar Survey
(Homeier et al. 1998; Homeier 2002, private communication)
and are noted in the tables.
Of these 44 WDs, 14 have spectra taken here, and the spectral types
agree very well.  (The star WD1352+004 was classified DB4, but here
is found to have Balmer absorption and is called DBA4.)
Most of the remaining stars can be identified in the imaging data,
but did not get spectroscopic fibers assigned.
Our spectroscopic identification of 269 WDs in this region is an
increase by a factor of more than six over the WDs known previously,
and gives some indication of the returns that can be expected from
the complete survey.
Our ``complete'' region has 10 square degrees of overlap with the
15-hour field of Evans (1992).  Evans finds 12 WDs in this overlap
region from his reduced proper motions.  We have spectra taken of five
(3 DB, 1 DA, 1 DA+M), and we have spectra of an additional
12 WDs that Evans did not find --- 10 with their proper motions below
his lower limit and 2 apparently too faint.  Again, SDSS is yielding
greater returns in the WDs identified.

With the WDs and hot subdwarfs in Tables 1--3 inside the 41-plate
area (omitting the stars in Table 2 outside this area),
we can look at the density on the sky of these hot stars.
Let us consider two subsamples:  hot stars with $(u-g)_0<0.0$ and
$(g-r)_0<-0.2$, and medium-hot stars with $0.0<(u-g)_0<0.6$ and
$(g-r)_0<-0.2$.  These subsamples need to be considered separately
because a higher priority is given for fiber assignment to
the ``HOT\_STANDARD'' stars with the bluer colors.
These two groups contain DA WDs with roughly $T>22000$
and $12000<T<22000$~K, respectively.  Taking stars with $15<g<20$,
and applying the full reddening corrections (which is not correct
for some of the brightest and reddest nearby WDs, but which will
not affect the conclusions), a summary is given in Tables 4 and 5.
In these tables, columns 2 and 3 give the number of stars with
observed spectra, and columns 4 and 5 give the corresponding
density on the sky.  However, SDSS only targets some fraction of
all hot stars for spectroscopic observations; the fraction can
be estimated by counting objects in the SDSS imaging catalogs
with stellar image profiles and with the same magnitude and
color limits as above.  The blue stars without spectra are assumed
to have the same DA/DB/sdB distributions found from the stars with
spectra in each color subsample at each magnitude.  Then the densities
in columns 4 and 5 can be adjusted to give the total sky densities.
The last columns in Tables 4 and 5 give these results for the total
density on the sky of blue stars with $15<g<20$ adjusted for
incomplete selection for obtaining spectra.  The tables show that
for brighter, hot stars ($g<19$ and $(u-g)_0<0.0$),
most stars have SDSS spectra.  For fainter stars and for the
medium-hot subsample, the fraction with spectra is roughly one-third.

The density of WDs found here is consistent with simulations
(Fan 1999) of 5 deg$^{-1}$ for WDs of all temperatures with $15<g<20$
at b=60$^\circ$.  The hot sdB/sdO stars outnumber hot WDs at bright
magnitudes, as was also found in the PG Survey \cite{gre86}.
The numbers reverse at fainter magnitudes, however,
as the space density of sdB stars drops at large distances
from the Galactic plane.  The reversal can be seen in
Table 5, where the numbers of WDs and hot subdwarfs are shown
from the hot-star subsample with $(u-g)_0<0.0$ and $(g-r)_0<-0.2$.
The star densities in Tables 4 and 5 will vary with Galactic
latitude --- the latitude dependence will be explored when
larger SDSS samples are available.

At cooler temperatures (6000--10000~K), WDs are still separated
in color from main-sequence stars, and many can be identified
using a combination of color and proper-motion selection.
For yet-cooler temperatures (4000--6000~K), proper motion will
be essential to isolate WDs, and even then contamination from
metal-poor halo subdwarf stars will be a severe difficulty.
Finally, WDs at temperatures below 4000~K have their red-IR flux
depressed by molecular hydrogen absorption (collision induced
absorption).  Their SDSS colors then become so distorted that they
are easily identified.  Stars SDSSJ1337+0001 and LHS3250
(Harris et al.~2001) are seen to be prominant outliers in Fig.~1b
and Table 2 --- they are still the only stars of this type
discovered or recovered in the SDSS.
Future papers will discuss these rare stars further as more are found.

\acknowledgments
Funding for the Sloan Digital Sky Survey (SDSS) has been provided by
the Alfred P. Sloan Foundation, the Participating Institutions,
the National Aeronautics and Space Administration, the National Science
Foundation, the U.S. Department of Energy, the Japanese
Monbukagakusho, and the Max Planck Society.
The SDSS is a joint project of The University of Chicago, Fermilab,
the Institute for Advanced Study, the Japan Participation Group,
The Johns Hopkins University, Los Alamos National Laboratory,
the Max-Planck-Institute for Astronomy (MPIA), the Max-Planck-Institute
for Astrophysics (MPA), New Mexico State University, University
of Pittsburgh, Princeton University, the United States Naval Observatory,
and the University of Washington.
We are grateful to P. Bergeron for calculating SDSS colors
of WD models, and D. Homeier for providing lists of WDs from the
Hamburg Survey in advance of publication.
SLH, NMS, and PS acknowledge support from NSF grant AST~0205875.
Studies of strongly magnetic stars and stellar systems at
Steward Observatory are supported by NSF grant AST 97-30792 to GS.

\clearpage

\figcaption[harris_fig1.ps]{
Color-color diagrams ($u-g$/$g-r$ in panel (a), $g-r$/$r-i$ in panel (b))
showing the WDs and hot subdwarfs in Tables 1--3.  Different types are
shown with different symbols.  Small dots show all objects with
stellar image profiles with $15 < g < 20$ in a region of 25~deg$^2$.
The curves show the colors of WD model atmospheres (\cite{ber95})
of pure H (solid curves) and pure He (dashed curves) with
log~$g = 7$, 8, and 9, where the log~$g = 7$ curve is toward the lower
right and the log~$g = 9$ curve is toward the upper left in panel (a).
The dotted lines with labels connect models
with the same effective temperature.
\label{fig1}}

\figcaption[harris_fig2.ps]{
Sequence of DA4 stars covering a range of magnitudes.
The $g$ magnitude is given at the left below each spectrum.
The coordinates (and in parentheses, the plate--fiber number of the
spectrum) are given at the right below each spectrum.
The spectra in this and following figures have been smoothed by 5 pixels
to a resolution of about 1200.
\label{fig2}}

\figcaption[da.ps]{
A sample of DA WDs with hydrogen atmospheres, arranged in order
of $u-i$ color and spectral type.  DA stars cooler than DA7
(7000~K) exist, but are not targeted for SDSS spectra because
their colors overlap with the stellar locus.
\label{fig3}}

\figcaption[db.ps]{
A sample of hot WDs with helium atmospheres, arranged in order
of $u-i$ color.  The first two are DO and the last five are DB,
except the sixth has hydrogen (weak H$\beta$, stronger H$\alpha$)
and is DBA.
\label{fig4}}

\figcaption[magda.ps]{
Twelve magnetic DAH stars, arranged in order of increasing field strength.
\label{fig5}}

\figcaption[magdb.ps]{
Three magnetic DBH stars.
\label{fig6}}

\figcaption[dqwk.ps]{
Ten DQ stars with weak carbon features, plotted in order of $u-i$ color.
The first four stars show the atomic lines of CI marked by dots at the
top of the plot.  The fourth and following stars show the Swan bands
of molecular carbon.  The heads of the three strongest (1,0), (0,0),
and (0,1) bands are marked by dashed lines.
\label{fig7}}

\figcaption[dqstr.ps]{
Nine DQ stars with strong Swan bands of molecular carbon,
plotted in order of $u-i$ color.  The heads of the three strongest
bands are marked by dashed lines.
\label{fig8}}

\figcaption[dzdc.ps]{
Sequence of all DZ stars and the brighter DC stars in the 41-plate
sample, plotted in order of $g-i$ color.
The top spectrum shows a DBZ star with He and Ca.
The remaining eight spectra show four typical DZ stars, with
Ca but no other elements detected,
and four DC stars with no definite features.
\label{fig9}}

\figcaption[dzstrong.ps]{
Sequence of ten DZ stars with Mg and/or Na lines present,
in addition to Ca.  They are plotted in order of $r-z$ color.
\label{fig10}}

\figcaption[dzfelines.ps]{
Blowup of four DZ spectra from Fig. 10 showing a rich set of
metallic lines.  The wavelengths of some of the detected lines
are marked at the top by open circles (Ca and Mg lines) and small
filled circles (Fe lines).
\label{fig11}}

\figcaption[specsdb.eps]{
The brightest of the hot subdwarfs with $g<16.7$, arranged by
$u-i$ color with bluer stars at the top.
\label{fig12}}

\clearpage
\end{document}